\documentclass[twocolumn,superscriptaddress,preprintnumbers,amsmath,amssymb,prd]{revtex4}
\usepackage{graphicx}
\UseRawInputEncoding

\begin{document}

\thispagestyle{empty}

\title{Comment on ``Revisiting the divergent multipole expansion of atom-surface interactions:
Hydrogen and positronium, $\alpha$-quartz, and physisorption" (arXiv:2308.04656v3)}

\author{
G.L.Klimchitskaya}
\affiliation{Central Astronomical Observatory at Pulkovo of the
Russian Academy of Sciences, Saint Petersburg,
196140, Russia}
\affiliation{Peter the Great Saint Petersburg
Polytechnic University, Saint Petersburg, 195251, Russia}

\begin{abstract}
Recently U. D. Jentschura [Phys. Rev. A $\bf 109$, 012802 (2024)] rederived the multipole
corrections to the dipole part of the atom-wall interaction described by the Lifshitz theory
using the concept of volume dielectric permittivity.
These corrections were computed for the hydrogen and positronium atoms in close proximity
to the $\alpha$-quartz wall and claimed to be numerically significant within the short-range
regime. Here, it is shown that the application areas of the obtained expressions both in the short-
and long-range  asymptotic regimes are indicated incorrectly, in contradiction with those
dictated by the Lifshitz theory. As a result, within the valid
application areas, all multipole corrections to the Casimir-Polder dipole part of the atom-wall
interaction turned out to be negligibly small.
\end{abstract}

\maketitle

Reference \cite{1} recalculated the higher-order multipole corrections (quadrupole, octupole,
and hexadecupole) to the Casimir-Polder dipole part of the atom-wall interaction energy
described by the Lifshitz formula \cite{2,3}. According to Ref. \cite{1}, the obtained quadrupole
correction is numerically significant for atom-surface interaction and modifies the van der Waals
corrected values of adsorption energy obtained in the literature \cite{4}.

Below we show that the application areas of the short- and long-range regimes for the
Casimir-Polder dipole part of atom-surface interaction and higher-order corrections to it are
indicated in Ref. \cite{1} incorrectly, in contradiction with the valid results established by the
founders of the Lifshitz theory. As a consequence, the quadrupole and all other higher-order
multipole corrections of Ref. \cite{1}, when computed in the proper areas of their validity, turn
out to be negligibly small, as compared to the dipole part given by the Lifshitz formula.

The dipole part of the interaction energy ${\cal E}_1(z)$ between an atom and a wall separated by
a distance $z$ is given by the Lifshitz formula (Eqs. (24) and (25) of Ref. \cite{1}). As to the
quadrupole,  ${\cal E}_2(z)$, octupole,  ${\cal E}_3(z)$, and hexadecupole,  ${\cal E}_4(z)$,
corrections to it, they are given by Eqs. (34), (40), and (46) of Ref. \cite{1}, respectively. All
the quantities  ${\cal E}_l(z)$ with $l=1$, 2, 3, and 4 are presented in the form of integrals
along the imaginary frequency axis, where the integrands depend on the dipole, octupole,
quadrupole, and hexadecupole atomic polarizabilities ${\alpha}_l(i\omega)$ and the dielectric
permittivity of wall material ${\varepsilon}(i\omega)$.

According to Ref. \cite{1}, the short-range distance regime of the interaction energies ${\cal E}_l(z)$
holds at separations satisfying the conditions
\begin{equation}
a_0 \ll z \ll \frac{a_0}{\alpha},
\label{eq1}
\end{equation}
\noindent
where
\begin{equation}
1~ {\rm a.u. length} = a_0 = \frac{4\pi\varepsilon_0\hbar^2}{e^2m_e} \approx 0.529~ {\rm \AA}
\label{eq2}
\end{equation}
\noindent
is the Bohr radius, $\varepsilon_0$ is the permittivity of free space, $e$ and $m_e$ are the
charge and mass of an electron, and $\alpha = e^2/(4\pi\varepsilon_0\hbar c) \approx 1/137$
is the fine structure constant.

From Eq. (\ref{eq1}), one can conclude that, according to Ref. \cite{1}, the short-range regime
holds for $z \geq 5~\rm {\AA} = 0.5$ nm. In fact, by citing Ref. \cite{5}, Ref. \cite{1}
advocates that the short-range expressions are actually valid down to distance regions of a few
angstroms away from the surface. The long-range regime, as stated in Ref. \cite{1}, holds at
atom-wall distances
\begin{equation}
z \gg \frac{a_0}{\alpha}.
\label{eq3}
\end{equation}

The analytic expressions for ${\cal E}_l(z)$ obtained in Ref. \cite{1} in the short-range regime
are given by
\begin{equation}
{\cal E}_l(z) \approx - \frac{C_{(2l+1)0}}{z^{2l+1}},
\label{eq4}
\end{equation}
where
\begin{equation}
C_{(2l+1)0} = - \frac{\hbar}{32\pi^2\varepsilon_0}\frac{l+1}{l} \int_{0}^{\infty}{\alpha}_l(i{\omega})
\frac{{\varepsilon}(i{\omega}) - 1}{{\varepsilon}(i{\omega}) +1} d {\omega}.
\label{eq5}
\end{equation}

The values of $C_{(2l+1)0}$ were computed for the atoms of hydrogen and positronium and the
$\alpha$-quartz wall using the multipole polarizabilities of these atoms and the improved formula for the
dielectric permittivity of the wall material (see Tables II and III in Ref. \cite{1}). To illustrate an
importance of the quadrupole correction compared to the leading dipole term, Ref. \cite{1} presents
the ratio of ${\cal E}_2$ to ${\cal E}_1$ for a positronium atom at $z = 10$ a.u. = 0.529 nm
distance from the wall
\begin{equation}
\frac{{\cal E}_2^{({\rm Ps})}(z = 10~ {\rm a.u.)}}{{\cal E}_1^{({\rm Ps})}(z = 10 ~{\rm a.u.)}} =
\frac{C_{50}^{({\rm Ps})}}{C_{30}^{({\rm Ps})}\times 10^2} \approx 0.124.
\label{eq6}
\end{equation}
\noindent
For a hydrogen atom, Ref. \cite{1} arrives at
\begin{equation}
\frac{{\cal E}_2^{({\rm H})}(z = 10~ {\rm a.u.)}}{{\cal E}_1^{({\rm H})}(z = 10 ~{\rm a.u.)}} =
\frac{C_{50}^{({\rm H})}}{C_{30}^{({\rm H})}\times 10^2} \approx 0.0296.
\label{eq7}
\end{equation}

According to Ref. \cite{1}, the results obtained in the short-range and long-range limits of the Lifshitz
formula for the Casimir-Polder energy  ${\cal E}_1$ are consistent with those in Refs. \cite{6,7,8}.
The reader is not informed, however, that in Refs. \cite{6,7,8} both the short- and long-range limits
are understood in a completely different way than in Eqs. (\ref{eq1}) and (\ref{eq3}).

The point is that the Lifshitz theory is the semiclassical one. It describes the wall material as a
continuous medium by means of the c-function, frequency-dependent volume dielectric permittivity,
and only the fluctuating electromagnetic field is quantized. For this reason, already in the classical
Lifshitz paper \cite{9}, it is stated that ``We can however approach this problem in purely
macroscopic fashion (since the distance between the bodies is assumed to be large compared to
interatomic distances)." The same statement is repeated in the seminal paper \cite{6}: ``...one
may approach the problem from a completely different and purely macroscopic point of view, in
which the interacting bodies are considered as continuous media. This approach is valid because
the distance between the two surfaces, although small, is large compared to the interatomic
distances in the bodies." The classical textbook \cite{2}, when discussing the allowed distances
between the two plates or an atom and a plate in the Lifshitz theory, also states: ``...this distance
satisfying only the one condition that it is large compared with interatomic distances in the bodies."
This condition is necessary for a wall material to be described by the volume dielectric permittivity.

As to the upper bound of the short-range regime of the Lifshitz formula, Refs. \cite{2,3,6,7,8,9}
state that this regime extends to distances much shorter than the characteristic wavelength $\lambda_0$
in the absorption spectrum of the wall material. To summarize the above information, the originators
of the Lifshitz theory found that the short-range regime of the Casimir-Polder interaction energy
${\cal E}_1$ holds not under the conditions of Eq. (\ref{eq1}) used in Ref. \cite{1} but under
the inequalities
\begin{equation}
d \ll z \ll \lambda_0,
\label{eq8}
\end{equation}
\noindent
where $d$ is an interatomic distance in the wall material. The same inequalities determine the short
distance regime for all multipole corrections ${\cal E}_l$ to ${\cal E}_1$.

In contrast with Ref. \cite{1}, the long-range regime holds not under the condition (\ref{eq3}) but
at \cite{2,3,6,7,8,9}
\begin{equation}
\lambda_0 \ll z \ll \frac{\hbar c}{k_BT} \approx 7.6~ {\mu}{\rm m},
\label{eq9}
\end{equation}
\noindent
where $k_B$ is the Boltzmann constant, $T$ is the temperature and  the room temperature value
$T=300$K was used in the right-hand side of Eq.  (\ref{eq9}). At $T=0$  the long-range regime
extends from separations much larger than $\lambda_0$ to infinity.

The question arises what is the actual size of the quadrupole and higher-order multipole
corrections to the dipole atom-wall interaction given by the Lifshitz theory? To answer this question,
we consider the shortest separation distance $z_{\rm min}$ satisfying Eq. (\ref{eq8}), where these
corrections are the most pronounced. Keeping in mind that the lattice constants of $\alpha$-quartz
are approximately equal to 5~\AA, we put $d = 0.5$ nm and choose $z_{\rm min}$ equal to
4 nm = 75.61 a.u., i.e., by a factor of 8 larger than $d$.

Using the data of Table III in Ref. \cite{1}, we calculate the ratio of the quadrupole correction,
${\cal E}_2$, to the main dipole part, ${\cal E}_1$, for the atom of positronium separated by the
distance $z_{\rm min}$ from the wall made of $\alpha$-quartz
\begin{equation}
\frac{{\cal E}_2^{({\rm Ps})}(z = 75.61~ {\rm a.u.)}}{{\cal E}_1^{({\rm Ps})}(z = 75.61 ~{\rm a.u.)}} =
\frac{C_{50}^{({\rm Ps})}}{C_{30}^{({\rm Ps})}\times (75.61)^2} \approx 2.2 \times 10^{-3}.
\label{eq10}
\end{equation}

In a similar way, using the data of Table II in Ref. \cite{1}, for a hydrogen atom spaced at the same
distance from the $\alpha$-quartz wall one obtains
\begin{equation}
\frac{{\cal E}_2^{({\rm H})}(z = 75.61~ {\rm a.u.)}}{{\cal E}_1^{({\rm H})}(z = 75.61 ~{\rm a.u.)}} =
\frac{C_{50}^{({\rm H})}}{C_{30}^{({\rm H})}\times (75.61)^2} \approx 5.2 \times 10^{-4}.
\label{eq11}
\end{equation}

Needless to say that all the higher-order multipole corrections are all the more insignificant. For
instance, the ratio of ${\cal E}_3^{({\rm H})}$ to the main part ${\cal E}_1^{({\rm H})}$ at
the same minimum distance satisfying Eq. (\ref{eq8}) is given by
\begin{equation}
\frac{{\cal E}_3^{({\rm H})}(z = 75.61~ {\rm a.u.)}}{{\cal E}_1^{({\rm H})}(z = 75.61 ~{\rm a.u.)}} =
\frac{C_{70}^{({\rm H})}}{C_{30}^{({\rm H})}\times (75.61)^4} \approx 7.56 \times 10^{-7}.
\label{eq12}
\end{equation}

It is necessary also to take into account that the optical data for a wall material used for
calculating the dielectric permittivity along the imaginary frequency axis are measured with some
error. This leads to approximately 0.5\% error in the calculated values of $C_{(2l + 1)0}$ \cite{3}.
One can conclude that, within the actual application area of the theory used for their calculation,
all the higher-order multipole corrections to the main dipole part of the atom-wall interaction are
negligibly small and their account is superfluous.

However, in order to support the application area (\ref{eq1}) of the short-distance regime,
Ref. \cite{1} remarks that at $z = 0.5$ nm $\approx 10$ a.u. the overlap between the wave functions
of the ground-state hydrogen atom and the wall atoms is already negligibly small. In addition,
Ref. \cite{1} mentions the result \cite{5} that at atom-wall separations down to 4 -- 7 a.u. the
interaction energy can be presented in the form
\begin{equation}
V(z) \approx -\frac{C_3}{(z - z_0)^3},
\label{eq13}
\end{equation}
\noindent
where $C_3 = C_{(3)0}$ is defined in Eq. (\ref{eq5}) with $l = 1$ and $z_0$ determines the position
of the so-called reference plane locating the atom-wall separation. The value of $z_0$ in Eq.  (\ref{eq13})
was obtained from the sum of the Casimir-Polder energy ${\cal E}_1$ and the phenomenological
correction to it of the form --$C_4/z^4$, which describes the lateral average potential with respect
to positions of particles on the surface \cite{4}. In so doing, the second-order perturbation theory
using the jellium model or the model of spherically symmetric nonoverlapping atoms was employed.

Both these remarks are in fact irrelevant to the application area (\ref{eq8}) of the Lifshitz formula
describing the atom-wall interaction, which makes it possible to consider the wall surface as a spatially
homogeneous plane. Note that at the separation distances of 4 -- 7 a.u. characteristic for a
physisorption Ref. \cite{5} also underlines the necessity for ``a good treatment of the spatial variation
of the interaction potential along the surface", which is beyond the Lifshitz theory, and considers it
as a problem to be solved in future. It should be taken into account that, although at such separations an
overlap of the wave functions might be small, the adhesive forces play an important role. In any
case, at separations below several interatomic distances $d$, the interaction energy should
be calculated not semiclassically, but using, e.g., the methods of density-functional theory \cite{10}.
This conclusion remains unchanged when considering the generalization
of the Lifshitz theory, which admits the spatially nonlocal dielectric
permittivities. Although such permittivities take into account spatial
information of the surface material, they can be introduced for only
the model of continuous medium, i.e., the separation from the surface
should be large enough to neglect the atomic structure.
Thus, although Ref. [5] cannot be considered as an extension of the
Lifshitz theory, it presents useful semiphenomenological method
for a description of physisorption.

It is common knowledge that in the descriptions of complicated physical
phenomena, such as physisorption, it is sometimes necessary to use the
computational results obtained using different theoretical approaches
outside of their application regions by joining them smoothly in the
intermediate transition regions. Just this is the case when the results
obtained by means of the density functional theory are combined with
computations using the Lifshitz theory. This does not mean, however, as
erroneously made in Ref. \cite{1}, that the physically justified
application region of the fundamental Lifshitz theory should be revised.

In addition to the interaction of an atom with a dielectric wall, Ref. \cite{1} considers the dipole part
of the atom-wall interaction for a perfectly conducting wall and the higher-order multipole
corrections to it (see Eqs. (20)-(23) of Ref. \cite{1}). It is stated that the short- and long-range
regimes are again given by Eqs. (\ref{eq1}) and (\ref{eq3}), respectively. According to Ref. \cite{1},
the short-range expressions for ${\cal E}_l^{({\rm H})}(z)$ and ${\cal E}_l^{({\rm Ps})}(z)$
(Eqs. (54) and (55) of Ref. \cite{1}) derived for a perfectly conducting wall ``are actually valid
down to distance regions of a few angstroms away from the surface." This statement is, however,
misleading because the walls made even of good conductors can be modelled by the perfect
conductor only at atom-wall separations exceeding approximately 1~$\mu$m \cite{3}.
Therefore, the results of Ref. \cite{1} obtained for a perfectly conducting wall in the limit of
short separations are irrelevant to any physical system.

To conclude, the higher-order multipole corrections to the Casimir-Polder dipole part of
the atom-wall interaction given by the Lifshitz theory are negligibly small. Relatively large
values of the quadrupole correction indicated in Ref. \cite{1} are explained by the fact that the
obtained expressions were used outside the application area of the concept of volume dielectric
permittivity.

\section*{Acknowledgments}
This commentary was supported by the State assignment for basic research
(project FSEG-2023-0016).


\end{document}